\documentclass[12pt]{article}
\usepackage{graphicx}
\usepackage{amsmath}
\usepackage{amssymb}
\usepackage{caption2}
\begin{document}
\bibliographystyle{prsty}
\begin{center}
{\large {\bf \sc{   Structure of the $D_{s0}(2317)$
 and the strong coupling constant $g_{D_{s0} D K}$ with
  the light-cone QCD sum rules  }}} \\[2mm]
Z. G. Wang$^{1}$ \footnote{Corresponding author; E-mail,wangzgyiti@yahoo.com.cn.  }, S. L. Wan$^{2}$     \\
$^{1}$ Department of Physics, North China Electric Power University, Baoding 071003, P. R. China \\
$^{2}$ Department of Modern Physics, University of Science and Technology of China, Hefei 230026, P. R. China \\
\end{center}

\begin{abstract}
In this article, we take the point of view that the  charmed scalar
 meson $D_{s0}(2317)$ is the conventional $c\bar{s}$ meson and
 calculate the strong coupling constant $g_{D_{s0} D K}$ within the
framework of the light-cone QCD sum rules approach.  The numerical
values  for  the large scalar-$DK$ coupling constant $g_{D_{s0} D
K}$ support the hadronic dressing mechanism. Just like the scalar
mesons $f_0(980)$ and $a_0(980)$,  the $D_{s0}(2317)$ may have small
scalar $c\bar{s}$ kernel of the typical $c\bar{s}$ meson size. The
strong coupling to the hadronic channels (or the virtual mesons
loops) may result in  smaller mass than the conventional scalar
$c\bar{s}$ meson in the constituent quark models,   and enrich the
pure $c\bar{s}$ state  with other components.  The $D_{s0}(2317)$
may spend part (or most part) of its lifetime as virtual $ D K $
state.
\end{abstract}


\section{Introduction}
The constituent quark model provides a rather successful description
of the spectrum of the mesons in terms of quark-antiquark bound
states, which fit into the suitable multiplets reasonably well.
However, the scalar mesons below $2GeV$ present a remarkable
exception as the structures of those mesons are not unambiguously
determined yet \cite{Godfray}. The light scalar mesons are the
subject of an intense and continuous controversy in clarifying the
hadron spectroscopy, the more elusive things are the constituent
structures of the $f_0(980)$ and $a_0(980)$ mesons with almost the
degenerate masses. Furthermore, the  discovery of the two
strange-charmed mesons $D_{s0} (2317)$ and $D_{s1} (2460)$ with
spin-parity $0^+$ and $1^+$ respectively has triggered hot  debate
on their nature, under-structures and whether it is necessary to
introduce  the exotic  states \cite{exp03}. The mass of the
$D_{s0}(2317)$ is significantly lower than the values of the $0^+$
state mass  from the quark models   and lattice simulations
\cite{QuarkLattice}. The difficulties to identify the $D_{s0}(2317)$
and $D_{s1}(2460)$ states with the conventional $c\overline{s}$
mesons are rather similar to those appearing in the light scalar
mesons below $1GeV$. Those two states $D_{s0} (2317)$ and $D_{s1}
(2460)$ lie just below the $D K$ and $D^\ast K$ threshold
respectively,  which are analogous to the situation that the scalar
mesons  $a_0(980)$ and $f_0(980)$ lie just below the $K\bar{K}$
threshold and  couple strongly to the nearby channels. The mechanism
responsible for the low-mass charmed scalar meson may be the same as
the light scalar nonet mesons, the $f_{0}(600)$, $f_{0}(980)$,
$a_{0}(980)$  and $K^{\ast}_{0}(800)$
\cite{ColangeloWang,ReviewScalar,WangScalar05}. There have been a
lot of explanations for their nature, for example, conventional
$c\bar{s}$ states \cite{2quark,Colangelo2005,Mehen04}, two-meson
molecular state \cite{2meson}, four-quark states \cite{4quark}, etc.
If we take the scalar mesons  $a_0(980)$ and $f_0(980)$ as four
quark states with the
 constituents  of scalar diquark-antidiquark  sub-structures, the
masses of the scalar nonet mesons below  $1GeV$ can be naturally
explained \cite{ReviewScalar,WangScalar05}.

 There are other possibilities beside the four-quark state explanations, for
example, the scalar mesons $a_0(980)$, $f_0(980)$ and $D_{s0}(2317)$
may have bare $q\overline{q}$ and $c\bar{s}$ kernels in the $P-$wave
states  with strong coupling  to the nearby threshold respectively,
the $S-$wave  virtual intermediate hadronic states (or the virtual
mesons loops) play a crucial role in the composition of those bound
states (or resonances due to the masses below or above the
thresholds). The hadronic dressing mechanism (or unitarized quark
models) takes the point of view that the $f_0(980)$, $a_0(980)$ and
$D_{s0}(2317)$ mesons have small $q\bar{q}$ and $c\bar{s}$  kernels
of the typical $q\bar{q}$ and $c\bar{s}$  mesons size respectively.
The strong couplings to the virtual intermediate hadronic states (or
the virtual mesons loops) may result in smaller masses than the
conventional scalar $q\bar{q}$ and $c\bar{s}$ mesons in the
constituent quark models, enrich the pure $q\bar{q}$ and $c\bar{s}$
states with other components\cite{HDress,UQM}. Those mesons may
spend part (or most part) of their lifetime as virtual $ K \bar{K} $
and $DK$ states \cite{HDress,UQM}. Despite what constituents they
may have, we have the fact that  they
 lie just a little below the $K\bar{K}$ and $DK$ threshold respectively, the
strong interactions with the $K\bar{K}$ and $DK$ thresholds will
significantly influence their dynamics, although the decay
$D_{s0}(2317)\rightarrow DK$ is kinematically suppressed.  It is
interesting to investigate the possibility of the hadronic dressing
mechanism.

In this article, we take the point of view that the scalar mesons
$f_0(980)$, $a_0(980)$ and $D_{s0}(2317)$  are the conventional
$q\bar{q}$ and $c\bar{s}$ state respectively, and calculate the
values of the strong coupling constant $g_{D_{s0} DK}$ within the
framework of the light-cone QCD sum rules approach. The light-cone
QCD sum rules approach carries out the operator product expansion
near the light-cone $x^2\approx 0$ instead of the short distance
$x\approx 0$ while the nonperturbative matrix elements  are
parameterized by the light-cone distribution amplitudes
 which classified according to their twists  instead of
 the vacuum condensates \cite{LCSR,LCSRreview}.

The article is arranged as: in Section 2, we derive the strong
coupling constant  $g_{D_{s0} D K}$ within the framework of the
light-cone QCD sum rules approach; in Section 3, the numerical
results and discussions; and in Section 4, conclusion.

\section{Strong coupling constant  $g_{D_{s0} D K}$  with light-cone QCD sum rules}

In the following, we write down the definition  for the strong
coupling constant $g_{D_{s0} D K}$ ,
\begin{eqnarray}
\langle K(q) D(p)|D_{s0}(p+q) \rangle=g_{D_{s0} D K} \, \, .
\end{eqnarray}
We study the strong coupling constant $g_{D_{s0} D K}$ with the
scalar interpolating current $J_{D_{s0}}(x)$  and choose the
 two-point correlation function $T_{\mu}(p,q)$,
\begin{eqnarray}
 T_{\mu}(p,q)&=&i \int d^4x \, e^{i p \cdot x} \,
\langle{K(q)}|T\left\{J^D_\mu(x) J_{D_{s0}}(0)\right\}|0\rangle \, , \\
J_{D_{s0}}(x)&=& \bar{c}(x)s(x) \, , \\
J^D_\mu(x)&=&{\bar u}(x)\gamma_\mu \gamma_5 c(x)\, .
\end{eqnarray}
Here the axial-vector current $J^D_\mu(x)$  interpolates the
pseudoscalar $D$ meson, and the external $K$ state has four momentum
$q$ with $q^2=M_K^2$. The correlation function $T_{\mu}(p,q)$ can be
decomposed as
\begin{eqnarray}
T_{\mu}(p,q)&=&T_{p}\left(p^2,(p+q)^2\right)p_{\mu}+T_{q}
\left(p^2,(p+q)^2\right)q_{\mu},
\end{eqnarray}
due to the tensor analysis.

According to the basic assumption of current-hadron duality in the
QCD sum rules approach \cite{SVZ79}, we can insert  a complete
series of intermediate states with the same quantum numbers as the
current operators $J_{D_{s0}}(x)$ and $J^D_\mu(x)$  into the
correlation function $T_{\mu}(p,q)$ to obtain the hadronic
representation. After isolating the ground state contributions from
the pole terms of the $D_{s0}(2317)$ and $D$  mesons, we get the
following result,
\begin{eqnarray}
T_p\left(p^2,(p+q)^2\right)p_\mu&=&\frac{<0\mid J^D_{\mu}\mid
D(p)><DK\mid D_{s0}> <D_{s0}(p+q)|J_{D_{s0}}\mid 0>}
  {\left(M_D^2-p^2\right)\left(M_{D_{s0}}^2-(p+q)^2\right)} + \cdots \nonumber \\
&=&\frac{i g_{D_{s0}DK} f_{D}f_{D_{s0}}M_{D_{s0}}  p_\mu}
  {\left(M_D^2-p^2\right)\left(M_{D_{s0}}^2-(p+q)^2\right)} + \cdots
  ,
\end{eqnarray}
where the following definitions have been used,
\begin{eqnarray}
<D_{s0}(p+q)\mid J_{f_0}(0)\mid 0>&=&f_{D_{s0}}M_{D_{s0}}\,, \nonumber\\
<0\mid J^D_{\mu}(0)\mid D(p)>&=&if_Dp_\mu~~.
\end{eqnarray}
Here we have not shown the contributions from the high resonances
and  continuum states explicitly as they are suppressed due to the
double Borel transformation. In the ground state approximation, the
tensor structure  $T_q\left(p^2,(p+q)^2\right) q_{\mu}$  has no
contributions and neglected.

In the following, we briefly outline the  operator product expansion
for the correlation function $T_{\mu}(p,q)$ in perturbative QCD
theory. The calculations are performed at the large space-like
momentum regions $(p+q)^2\ll 0$  and  $p^2\ll 0$, which correspond
to the small light-cone distance $x^2\approx 0$ required by the
validity of the operator product expansion approach. We write down
the propagator of a massive quark in the external gluon field in the
Fock-Schwinger gauge firstly \cite{Belyaev94},
\begin{eqnarray}
&&\langle 0 | T \{q_i(x_1)\, \bar{q}_j(x_2)\}| 0 \rangle =
 i \int\frac{d^4k}{(2\pi)^4}e^{-ik(x_1-x_2)}\nonumber\\
 &&\left\{
\frac{\not\!k +m}{k^2-m^2} \delta_{ij} -\int\limits_0^1 dv\, g_s \,
G^{\mu\nu}_a(vx_1+(1-v)x_2)
\left (\frac{\lambda^a}{2} \right )_{ij} \right. \nonumber \\
&&\left. \Big[ \frac12 \frac {\not\!k
+m}{(k^2-m^2)^2}\sigma_{\mu\nu} - \frac1{k^2-m^2}v(x_1-x_2)_\mu
\gamma_\nu \Big]\right\}\, ,
\end{eqnarray}
here $G^{\mu \nu }_a$ is the gluonic field strength, $g_s$ denotes
the strong coupling constant. Substituting the above $c$  quark
propagator and the corresponding $K$ meson light-cone distribution
amplitudes into the correlation function $T_{\mu}(p,q)$ in Eq.(2)
and completing the integrals over the variables  $x$ and $k$,
finally we obtain the result,
\begin{eqnarray}
&&T_p(p^2,(p+q)^2)=\nonumber\\
&&if_K \int_0^1 du\left\{ {M_K^2 \over m_s} \varphi_p (u) {1 \over
m_c^2-(p+uq)^2}-2\left[ m_c g_2(u)+{M_K^2 \over 6 m_s}
\varphi_\sigma(u)(p \cdot q+u M_K^2) \right] \right.\nonumber \\
&&\left.{1 \over [m_c^2-(p+uq)^2]^2} \right\} +if_{3K} M_K^2
\int_0^1 dv \left(2v-3 \right) \int {\cal D}\alpha_i
\varphi_{3K}(\alpha_i) \nonumber\\
&&{1 \over \{ m_c^2-[p+q(\alpha_1+v\alpha_3)]^2 \}^2 }   -4 if_K m_c
M_K^2 \left\{ \int_0^1 dv (v-1) \int_0^1 d \alpha_3
\int_0^{\alpha_3} d \beta  \right.
 \nonumber \\
&&\left.\int_0^{1-\beta}d
\alpha{\Phi(\alpha,1-\alpha-\beta,\beta)\over
\{[p+(1-\alpha_3+v\alpha_3)q]^2-m_c^2\}^3}+
 \int_0^1 dv  \int_0^1 d \alpha_3 \int_0^{1-\alpha_3} d
\alpha_1\int_0^{\alpha_1}d \alpha \right.\nonumber\\
&&\left.{\Phi(\alpha,1-\alpha-\alpha_3,\alpha_3) \over
\{[p+(\alpha_1+v \alpha_3)q]^2-m_c^2\}^3 } \right\} .
\end{eqnarray}
In calculation, the following  two-particle and three-particle $K$
meson light-cone distribution amplitudes are useful
\cite{LCSR,LCSRreview,Belyaev94,Ball98,Ball06},
\begin{eqnarray}
&&<K(q)| {\bar u} (x) \gamma_\mu \gamma_5 s(0) |0> = -i f_K q_\mu
\int_0^1 du \; e^{i u q \cdot x} [\varphi_K(u)+x^2
g_1(u)]\nonumber\\
&&+f_K \left(x_\mu - {q_\mu x^2 \over q \cdot x}\right)
\int_0^1 du \; e^{i u q \cdot x} g_2(u)  , \nonumber\\
&&<K(q)| {\bar u} (x) i \gamma_5 s(0) |0> = {f_K M_K^2 \over m_s}
\int_0^1 du \; e^{i u q \cdot x} \varphi_p(u)  \hskip 3 pt ,  \nonumber\\
&&<K(q)| {\bar u} (x) \sigma_{\mu \nu} \gamma_5 s(0) |0> =i(q_\mu
x_\nu-q_\nu x_\mu)  {f_K M_K^2 \over 6 m_s} \int_0^1 du \;
e^{i u q \cdot x} \varphi_\sigma(u),  \nonumber\\
&&<K(q)| {\bar u} (x) \sigma_{\alpha \beta} \gamma_5 g_s G_{\mu
\nu}(v x)s(0) |0>=\nonumber\\
&&i f_{3 K}[(q_\mu q_\alpha g_{\nu
\beta}-q_\nu q_\alpha g_{\mu \beta}) -(q_\mu q_\beta g_{\nu
\alpha}-q_\nu q_\beta g_{\mu \alpha})] \int {\cal D}\alpha_i \;
\varphi_{3 K} (\alpha_i)
e^{iq \cdot x(\alpha_1+v \alpha_3)} ,\nonumber\\
&&<K(q)| {\bar u} (x) \gamma_{\mu} \gamma_5 g_s G_{\alpha
\beta}(vx)s(0) |0>=\nonumber\\
&&f_K \Big[ q_{\beta} \Big( g_{\alpha \mu}-{x_{\alpha}q_{\mu} \over
q \cdot x} \Big) -q_{\alpha} \Big( g_{\beta \mu}-{x_{\beta}q_{\mu}
\over q \cdot x} \Big) \Big] \int {\cal{D}} \alpha_i
\varphi_{\bot}(\alpha_i)
e^{iq \cdot x(\alpha_1 +v \alpha_3)}\nonumber \\
&&+f_K {q_{\mu} \over q \cdot x } (q_{\alpha} x_{\beta}-q_{\beta}
x_{\alpha}) \int {\cal{D}} \alpha_i \varphi_{\|} (\alpha_i)
e^{iq \cdot x(\alpha_1 +v \alpha_3)} \hskip 3 pt ,  \nonumber\\
&&<K(q)| {\bar u} (x) \gamma_{\mu}  g_s \tilde G_{\alpha
\beta}(vx)s(0) |0>=\nonumber\\
& &i f_K \Big[ q_{\beta} \Big( g_{\alpha \mu}-{x_{\alpha}q_{\mu}
\over q \cdot x} \Big) -q_{\alpha} \Big( g_{\beta
\mu}-{x_{\beta}q_{\mu} \over q \cdot x} \Big) \Big] \int {\cal{D}}
\alpha_i \tilde \varphi_{\bot}(\alpha_i)
e^{iq\cdot x(\alpha_1 +v \alpha_3)}\nonumber \\
&&+i f_K {q_{\mu} \over q \cdot x } (q_{\alpha}
x_{\beta}-q_{\beta} x_{\alpha}) \int {\cal{D}} \alpha_i \tilde
\varphi_{\|} (\alpha_i) e^{iq \cdot x(\alpha_1 +v \alpha_3)} \hskip 3 pt
.
\end{eqnarray}
Here the operator $\tilde G_{\alpha \beta}$  is the dual of the
$G_{\alpha \beta}$, $\tilde G_{\alpha \beta}= {1\over 2}
\epsilon_{\alpha \beta \delta \rho} G^{\delta \rho} $,
${\cal{D}}\alpha_i$ is defined as ${\cal{D}} \alpha_i =d \alpha_1 d
\alpha_2 d \alpha_3 \delta(1-\alpha_1 -\alpha_2 -\alpha_3)$ and
$\Phi(\alpha_1,\alpha_2,\alpha_3)=\varphi_{\bot}+\varphi_{\|}-\tilde
\varphi_{\bot}-\tilde \varphi_{\|}$. The twist-3 and twist-4
light-cone distribution amplitudes can be parameterized as
\begin{eqnarray}
\varphi_p(u,\mu)&=&1+\left(30\eta_3-\frac{5}{2}\rho^2\right)C_2^{\frac{1}{2}}(2u-1)\nonumber \\
&+&\left(-3\eta_3\omega_3-\frac{27}{20}\rho^2-\frac{81}{10}\rho^2\tilde{a}_2\right)C_4^{\frac{1}{2}}(2u-1),  \nonumber \\
\varphi_\sigma(u,\mu)&=&6u(1-u)\left(1
+\left(5\eta_3-\frac{1}{2}\eta_3\omega_3-\frac{7}{20}\rho^2-\frac{3}{5}\rho^2\tilde{a}_2\right)C_2^{\frac{3}{2}}(2u-1)\right), \nonumber \\
\phi_{3K}(\alpha_i,\mu) &=& 360 \alpha_1 \alpha_2 \alpha_3^2 \left
(1 +\lambda_3(\alpha_1-\alpha_2)+ \omega_3 \frac{1}{2} ( 7 \alpha_3
- 3) \right) , \nonumber\\
\phi_{\perp}(\alpha_i,\mu) &=& 30
\delta^2(\mu)(\alpha_1-\alpha_2)\alpha_3^2\left( \frac{1}{3} + 2
\epsilon (\mu) (1 - 2 \alpha_3) \right)  \, ,
\nonumber \\
\phi_{||}(\alpha_i,\mu) &=& 120 \delta^2(\mu) \epsilon (\mu)  (\alpha_1-\alpha_2) \alpha_1 \alpha_2 \alpha_3  \, ,
\nonumber\\
\tilde{\phi}_{\perp}(\alpha_i,\mu) &=& 30 \delta^2(\mu) \alpha_3^2 (
1 - \alpha_3) \left( \frac{1}{3} + 2 \epsilon (\mu)  (1 - 2
\alpha_3) \right) \, ,
 \nonumber\\
\tilde{\phi}_{||}(\alpha_i,\mu) &=& -120 \delta^2(\mu) \alpha_1
\alpha_2 \alpha_3 \left( \frac{1}{3} + \epsilon (\mu) (1 - 3
\alpha_3) \right) \, , \nonumber\\
g_2(u,\mu)&=&\frac{10}{3}\delta^2(\mu)u(1-u)(2u-1)\, ,
\end{eqnarray}
where  $ C_2^{\frac{1}{2}}$, $ C_4^{\frac{1}{2}}$
 and $ C_2^{\frac{3}{2}}$ are Gegenbauer polynomials,
 $\epsilon=\frac{21}{8}\omega_4$,
 $\eta_3=\frac{f_{3K}}{f_K}\frac{m_q+m_s}{M_K^2}$ and  $\rho^2={m_s^2\over M_K^2}$
 \cite{LCSR,LCSRreview,Belyaev94,Ball98,Ball06}.
The parameters in the light-cone distribution amplitudes can be
estimated from the QCD sum rules approach
\cite{LCSR,LCSRreview,Belyaev94,Ball98,Ball06}. In this article,
the energy scale $\mu$ is chosen to be  $\mu=1GeV$.

Now we perform the double Borel transformation with respect to  the
variables $Q_1^2=-p^2$ and  $Q_2^2=-(p+q)^2$ for the correlation
function $T_p(p^2,(p+q)^2)$ in Eq.(6),   and obtain the analytical
expression for the invariant function in the hadronic
representation,
\begin{eqnarray}
B_{M_2^2}B_{M_1^2}T_p(Q_1^2,Q_2^2)&=&i
g_{D_{s0}DK}f_{D}f_{D_{s0}}M_{D_{s0}}
e^{-M^2_D/M_1^2}e^{-M^2_{D_{s0}}/M_2^2} +\cdots,
\end{eqnarray}
here we have not shown  the contributions from the high resonances
and continuum states  explicitly for simplicity. In order to match
the duality regions below the thresholds $s_0$ and $s_0'$ for the
interpolating currents $J^{D}_{\mu}(x)$   and  $J_{D_{s0}}(x)$
respectively, we can express the correlation function $T_p$ at the
level of quark-gluon degrees of freedom into the following form,
\begin{eqnarray}
T_p(p^2,(p+q)^2)= i\int ds ds^\prime {\rho(s,s^\prime) \over (s-p^2)
[s^\prime-(p+q)^2]} \, ,
\end{eqnarray}
then we  perform the double Borel transformation with respect to the
variables $Q_1=-p^2$ and $Q_2^2=-(p+q)^2$  directly. However, the
analytical expressions for the spectral density $\rho(s,s')$ is hard
to obtain, we have to resort to  some approximations.  As the
contributions
 from the higher twist terms  are  suppressed by more powers of
 $\frac{1}{-p^2}$ or $\frac{1}{-(p+q)^2}$, the continuum subtractions will not affect the results remarkably,
here we will use the expressions in Eq.(9) for the three-particle
(quark-antiquark-gluon) twist-3 and twist-4  terms; in fact, their
contributions are of minor importance. The dominating contributions
come from the two-particle twist-3 terms involving the
$\varphi_p(u)$ and $\varphi_\sigma(u)$, we preform the same  trick
as Refs.\cite{Belyaev94,Kim} and expand the amplitudes
$\varphi_p(u)$ and $\varphi_\sigma(u)$ in terms of polynomials of
$1-u$,
\begin{eqnarray}
\varphi_p(u)+{d\varphi_\sigma(u)\over 6du}&=&\sum_{k=0}^N b_k(1-u)^k
\nonumber\\
 &=&b_k \left(\frac{s-m_c^2}{s-p^2}\right)^k,
\end{eqnarray}
then introduce the variable $s'$ and the spectral density is
obtained. After straightforward but cumbersome calculations, we can
obtain the final expression for the double Borel transformed
correlation function $T_p(M_1^2,M_2^2)$ at the level of quark-gluon
degrees of freedom. The masses of  the charmed mesons are
$M_{D_{s0}}=2.317GeV$ and $M_{D}=1.865GeV$,
$\frac{M_D}{M_D+M_{D_{s0}}}\approx0.45$,
 there exists an overlapping working window for the two Borel
parameters $M_1^2$ and $M_2^2$, it's convenient to take the value
$M_1^2=M_2^2$,  $u_0=\frac{M_1^2}{M_1^2+M_2^2}=\frac{1}{2}$,
$M^2=\frac{M_1^2M_2^2}{M_1^2+M_2^2}$,  furthermore, the $K$ meson
light-cone distribution amplitudes are known quite well at the value
$u_0=\frac {1}{2}$.  We can  introduce the threshold parameter $s_0$
and make the simple replacement,
\begin{eqnarray}
e^{-\frac{m_c^2+u_0(1-u_0)M_K^2}{M^2}} \rightarrow
e^{-\frac{m_c^2+u_0(1-u_0)M_K^2}{M^2} }-e^{-\frac{s_0}{M^2}}
\nonumber
\end{eqnarray}
 to subtract the contributions from the high resonances  and
  continuum states \cite{Belyaev94},
\begin{eqnarray}
&&B_{M_2^2}B_{M_1^2}T_p =\nonumber\\
&&i\left\{ {f_K M^2 M_K^2 \over m_s}
\left(e^{-\frac{m_c^2+u_0(1-u_0)M_K^2}{M^2}}-e^{-\frac{s_0}{M^2}}\right)\left(\varphi_p(u)+{d\varphi_\sigma(u)\over
6du}\right)
 \right. \nonumber \\
&&+e^{-\frac{m_c^2+u_0(1-u_0)M_K^2}{M^2}}\left[-2 f_K m_c  \,
g_2(u_0)\right.\nonumber\\
&&+f_{3K} M_K^2  \int_0^{u_0} d \alpha_1
\int_{u_0-\alpha_1}^{1-\alpha_1} {d \alpha_3 \over \alpha_3}
\varphi_{3K}(\alpha_1,1-\alpha_1-\alpha_3,\alpha_3)
\left(2{u_0-\alpha_1 \over \alpha_3}-3 \right) \nonumber \\
&&- { 2f_K m_c M_K^2\over M^2} (1-u_0) \int_{1-u_0}^1 {d \alpha_3
\over
\alpha_3^2}  \int_0^{\alpha_3}d \beta \int_0^{1-\beta}d \alpha \Phi(\alpha,1-\alpha-\beta,\beta)   \nonumber \\
&&+ {2f_Km_c M_K^2\over M^2}\left( \int_0^{1-u_0} {d \alpha_3 \over
\alpha_3} \int_{u_0-\alpha_3}^{u_0} d \alpha_1
\int_0^{\alpha_1}d\alpha+ \int_{1-u_0}^1 {d \alpha_3 \over \alpha_3}
\int_{u_0-\alpha_3}^{1-\alpha_3} d \alpha_1 \int_0^{\alpha_1}d\alpha
\right)\nonumber\\
&&\left.\left.\Phi(\alpha,1-\alpha-\alpha_3,\alpha_3) \right]
\right\} .
\end{eqnarray}
A slight different manipulation (with the techniques taken in the
Ref.\cite{ColangeloWang}) for the dominating contributions come from
the terms involving the two-particle twist-3 light-cone distribution
amplitudes $\varphi_p(u)$ and $\varphi_\sigma(u)$ leads to the
following result,
\begin{eqnarray}
&&B_{M_2^2}B_{M_1^2}T_p = \nonumber\\
&&ie^{-\frac{m_c^2+u_0(1-u_0)M_K^2}{M^2}} \left\{ {f_K M^2 M_K^2
\over m_s} \, \, \sum_{k=0}^N b_k \left({M^2 \over M_1^2}\right)^k
\left(1- e^{-\frac{s_0-m_c^2}{M^2}} \sum_{i=0}^k
{\left(\frac{s_0-m_c^2}{M^2}\right)^i \over i!} \right)
 \right. \nonumber \\
&&-2 f_K m_c   g_2(u_0)\nonumber\\
&&+f_{3K} M_K^2 \int_0^{u_0} d \alpha_1
\int_{u_0-\alpha_1}^{1-\alpha_1} {d \alpha_3 \over \alpha_3}
\varphi_{3K}(\alpha_1,1-\alpha_1-\alpha_3,\alpha_3)
\left(2{u_0-\alpha_1 \over \alpha_3}-3 \right)\nonumber \\
&&- { 2f_K m_c M_K^2\over M^2} (1-u_0) \int_{1-u_0}^1 {d \alpha_3
\over
\alpha_3^2}  \int_0^{\alpha_3}d \beta \int_0^{1-\beta}d \alpha \Phi(\alpha,1-\alpha-\beta,\beta)   \nonumber \\
&&+ {2f_Km_c M_K^2\over M^2}\left( \int_0^{1-u_0} {d \alpha_3 \over
\alpha_3} \int_{u_0-\alpha_3}^{u_0} d \alpha_1
\int_0^{\alpha_1}d\alpha+ \int_{1-u_0}^1 {d \alpha_3 \over \alpha_3}
\int_{u_0-\alpha_3}^{1-\alpha_3} d \alpha_1 \int_0^{\alpha_1}d\alpha
\right)\nonumber\\
&&\left.\Phi(\alpha,1-\alpha-\alpha_3,\alpha_3)  \right\} .
\end{eqnarray}
In deriving the above expressions for
$\varphi_p(u)+{d\varphi_\sigma(u)\over 6du}$, we have neglected the
terms $\sim M_K^4$, here $u_0=\frac{M_1^2}{M_1^2+M_2^2}$ and
$M^2=\frac{M_1^2M_2^2}{M_1^2+M_2^2}$.

Matching the Eq.(12) with the Eqs.(15-16) below the threshold $s_0$
and then we obtain two sum rules for the strong coupling constant
$g_{D_{s0}DK}$,
\begin{eqnarray}
&&g_{D_{s0}DK}=\nonumber\\
&&  \frac{1}{f_D
f_{D_{s0}}M_{D_{s0}}}e^{\frac{M^2_{D_{s0}}}{M_2^2}+\frac{M^2_D}{M_1^2}}
\left\{ {f_K M^2 M_K^2 \over m_s}
\left(e^{-\frac{m_c^2+u_0(1-u_0)M_K^2}{M^2}}-e^{-\frac{s_0}{M^2}}\right)\right.\nonumber\\
&& \left(\varphi_p(u)+{d\varphi_\sigma(u)\over 6du}\right)
+e^{-\frac{m_c^2+u_0(1-u_0)M_K^2}{M^2}}\left[-2 f_K m_c  \,
g_2(u_0)\right.\nonumber \\
&&+f_{3K} M_K^2  \int_0^{u_0} d \alpha_1
\int_{u_0-\alpha_1}^{1-\alpha_1} {d \alpha_3 \over \alpha_3}
\varphi_{3K}(\alpha_1,1-\alpha_1-\alpha_3,\alpha_3)
\left(2{u_0-\alpha_1 \over \alpha_3}-3 \right) \nonumber \\
&&- { 2f_K m_c M_K^2\over M^2} (1-u_0) \int_{1-u_0}^1 {d \alpha_3
\over
\alpha_3^2}  \int_0^{\alpha_3}d \beta \int_0^{1-\beta}d \alpha \Phi(\alpha,1-\alpha-\beta,\beta)  \nonumber \\
&&+ {2f_Km_c M_K^2\over M^2}\left( \int_0^{1-u_0} {d \alpha_3 \over
\alpha_3} \int_{u_0-\alpha_3}^{u_0} d \alpha_1
\int_0^{\alpha_1}d\alpha+ \int_{1-u_0}^1 {d \alpha_3 \over \alpha_3}
\int_{u_0-\alpha_3}^{1-\alpha_3} d \alpha_1 \int_0^{\alpha_1}d\alpha
\right)\nonumber\\
&&\left.\left.\Phi(\alpha,1-\alpha-\alpha_3,\alpha_3) \right]
\right\} \, \, ;
\end{eqnarray}

\begin{eqnarray}
&&g_{D_{s0}DK}= \nonumber\\
&&  \frac{1}{f_D
f_{D_{s0}}M_{D_{s0}}}e^{\frac{M^2_{D_{s0}}}{M_2^2}+\frac{M^2_D}{M_1^2}-\frac{m_c^2+u_0(1-u_0)M_K^2}{M^2}}
\left\{ {f_K M^2 M_K^2 \over m_s} \, \, \sum_{k=0}^N b_k \left({M^2
\over M_1^2}\right)^k \right.\nonumber\\
&&\left.\left(1- e^{-\frac{s_0-m_c^2}{M^2}} \sum_{i=0}^k
{\left(\frac{s_0-m_c^2}{M^2}\right)^i \over i!} \right)
 -2 f_K m_c  \, g_2(u_0)\right. \nonumber \\
&&+f_{3K} M_K^2  \int_0^{u_0} d \alpha_1
\int_{u_0-\alpha_1}^{1-\alpha_1} {d \alpha_3 \over \alpha_3}
\varphi_{3K}(\alpha_1,1-\alpha_1-\alpha_3,\alpha_3)
\left(2{u_0-\alpha_1 \over \alpha_3}-3 \right)\nonumber \\
&&- { 2f_K m_c M_K^2\over M^2} (1-u_0) \int_{1-u_0}^1 {d \alpha_3
\over
\alpha_3^2}  \int_0^{\alpha_3}d \beta \int_0^{1-\beta}d \alpha \Phi(\alpha,1-\alpha-\beta,\beta) \nonumber \\
&&+ {2f_Km_c M_K^2\over M^2}\left( \int_0^{1-u_0} {d \alpha_3 \over
\alpha_3} \int_{u_0-\alpha_3}^{u_0} d \alpha_1
\int_0^{\alpha_1}d\alpha+ \int_{1-u_0}^1 {d \alpha_3 \over \alpha_3}
\int_{u_0-\alpha_3}^{1-\alpha_3} d \alpha_1 \int_0^{\alpha_1}d\alpha
\right)\nonumber\\
&&\left.\Phi(\alpha,1-\alpha-\alpha_3,\alpha_3)  \right\} \,
\end{eqnarray}
corresponding to the Eq.(15) and Eq.(16) respectively.

\section{Numerical results and discussions}
The parameters are taken as $m_s=(140\pm 10 )MeV$, $m_c=(1.25\pm
0.10)GeV$, $\lambda_3=1.6\pm0.4$, $f_{3K}=(0.45\pm0.15)\times
10^{-2}GeV^2$, $\omega_3=-1.2\pm0.7$, $\delta^2=(0.20 \pm0.06
)GeV^2$, $\omega_4=0.2\pm0.1$, $\tilde a_2=0.25\pm 0.15$
\cite{LCSR,LCSRreview,Belyaev94,Ball98,Ball06}, $f_K=0.160GeV$,
$M_K=498MeV$, $M_{D_{s0}}(2317)=2.317GeV$, $M_D=1.865GeV$,
$f_{D}=(0.23\pm0.02)GeV$ \cite{WangD}, and
$f_{D_{s0}}=(0.225\pm0.025)GeV$ \cite{Colangelo2005}. The duality
thresholds $s_0$ in Eqs.(17-18) are taken as $s_0=(6.1-6.5)GeV^2$
 to avoid possible  contaminations from the high resonances and
continuum states, from the Fig.1, we can see that the numerical
results are not sensitive to the threshold  parameter $s_0$ in this
region. The Borel parameters are chosen as $10GeV^2 \le M_1^2=M_2^2
\le 20 GeV^2$ and $5GeV^2\le M^2 \le 10 GeV^2$, in those regions,
the values of  the strong coupling constant $g_{D_{s0} DK}$ are
rather stable from the sum rule in Eq.(17) with the simple
subtraction, which are shown  in the Figs.1-7. However, the values
from the sum rule in Eq.(18) with the more sophisticated subtraction
are not stable according to the variations of the Borel parameter
$M^2$.

The uncertainties of the five parameters $\delta^2$, $\omega_4$,
$\omega_3$, $\lambda_3$ and $\tilde{a}_2$ can  not result in large
uncertainties for the numerical values. The main uncertainties come
from the five parameters $f_{3K}$, $m_s$, $m_c$, $f_D$ and
$f_{D_{s0}}$, small variations of those parameters can lead to
relatively large changes for the numerical values, which are shown
in the Fig.2, Fig.3, Fig.4, Fig.5 and Fig.6, respectively. Taking
into account all the uncertainties,  finally we obtain the numerical
results for the strong coupling constant,
\begin{eqnarray}
 g_{D_{s0}DK} =(9.3^{+2.7}_{-2.1}) GeV  .
\end{eqnarray}
The strong coupling constant $g_{D_{s0}DK}$ can be related to the
parameter $h$ in the heavy-light Chiral perturbation theory
\cite{Casalbuoni97,Colangelo95},
\begin{eqnarray}
 g_{SP\pi} &=& \sqrt{M_{S} M_P} \frac{M_{S}^2-M_P^2}{M_{S}} \frac{|h|}{f_\pi} \,  , \nonumber \\
\end{eqnarray}
here the $S$ are the scalar heavy   mesons with $0^+$, the $P$ are
the heavy pseudoscalar  mesons with $0^-$, and  the $\pi$ stand for
 the light pseudoscalar mesons. The parameter $h$ has been estimated
with the light-cone QCD sum rules \cite{Colangelo95}, quark models
\cite{Becirevic99},  Adler-Weisberger type sum rule \cite{Chow96},
and extracted from the experimental data \cite{Mehen04}, the values
are listed in the Table.1, from those values we can estimate the
values of the corresponding strong coupling constant $g_{D_{s0}DK}$
in the $SU(3)$ limit for the light pseudoscalar mesons. The
dimensionless  effective coupling constant $\Gamma/k$ with the value
$\Gamma/k=0.46(9)$ from Lattice QCD \cite{UKQCD04}  is somewhat
smaller than the values extracted from the experimental data
$\Gamma/k=0.73^{+28}_{-24}$, here the $\Gamma$ is the decay width
and the $k$ is the decay momentum.
\begin{table}
\begin{center}
\begin{tabular}{c|c|c}
\hline\hline
      $|h|$   & $g_{D_{s0}DK}(GeV)$ &Reference  \\ \hline
       $0.52\pm0.17$ & $5.5\pm1.8$& \cite{Colangelo95}  \\ \hline
        0.536  & $5.68$&\cite{Becirevic99} \\ \hline
        $ <0.93$& $<9.86$&\cite{Chow96} \\ \hline
        $0.57-0.74$ &6.0-7.8&  \cite{Mehen04}\\ \hline
         &10.203&  \cite{Guo06}\\ \hline
       $0.88^{+0.26}_{-0.20}$  &$9.3^{+2.7}_{-2.1}$& This work \\ \hline  \hline
\end{tabular}
\end{center}
\caption{ Numerical values for the  parameter $h$,  and the
corresponding values for the strong coupling constant $g_{D_{s0}DK}$
in the $SU(3)$ limit. }
\end{table}
Our numerical values $g_{D_{s0}DK} =(9.3^{+2.7}_{-2.1}) GeV$ are
somewhat larger comparing with the existing estimations in
Refs.\cite{Colangelo95,Becirevic99,Chow96,Mehen04} and about four
times as large as the energy scale $M_{D_{s0}}=2.317GeV$, and favor
the hadronic dressing mechanism.

Here we will take a short digression to discuss the hadronic
dressing mechanism \cite{HDress,UQM}, one can consult the original
literatures  for the details. In the conventional constituent quark
models, the mesons are taken as quark-antiquark bound states. The
spectrum can be obtained by solving the corresponding Schrodinger's
or Dirac's equations with the phenomenological potential which
trying to incorporate the observed properties of the strong
interactions, such as the asymptotic freedom and confinement. The
solutions can be referred as confinement bound states or bare
quark-antiquark states (or kernels). If we switch on the hadronic
interactions between the confinement bound states and the free
ordinary  two-meson states, the situation becomes more complex. With
the increasing hadronic coupling constants, the contributions from
the hadronic loops of the intermediate mesons become larger and  the
bare quark-antiquark states can be distorted greatly. There may be
double poles or several poles   in the scattering amplitudes with
the same quantum number as the bare quark-antiquark kernels; some
ones stem  from the bare quark-antiquark kernels while the others
originate from the continuum states. The strong coupling may enrich
the bare quark-antiquark states with other components for example
virtual mesons pairs and  spend part (or most part) of their
lifetime as virtual mesons pairs.

 The large values for the strong coupling constant $g_{D_{s0}DK}$ obviously support
  the hadronic
dressing mechanism, the $D_{s0}(2317)$ (just like the scalar mesons
$f_0(980)$ and $a_0(980)$, see Ref.\cite{ColangeloWang}) can be
taken as having small scalar $c\bar{s}$ kernel of typical meson size
with large virtual S-wave $DK$ cloud. In Ref.\cite{Guo06}, the
authors analyze the unitarized two-meson scattering amplitudes from
the heavy-light Chiral Lagrangian,  and observe that the scalar
meson $D_{s0}(2317)$ appears as the bound state pole with the strong
coupling constant $g_{D_{s0}DK}=10.203GeV$. Our numerical results
$g_{D_{s0}DK} =(9.3^{+2.7}_{-2.1}) GeV$ are certainly reasonable and
can make robust predictions. However, we take the point of view that
the  scalar meson $D_{s0}(2317)$ be bound state  in the sense that
it appears below the $DK$ threshold, its constituents may be the
bare $c\bar{s}$ state, the virtual  $DK$ pair and their mixing,
rather than the $DK$ bound state.

\begin{figure}
 \centering
 \includegraphics[totalheight=7cm]{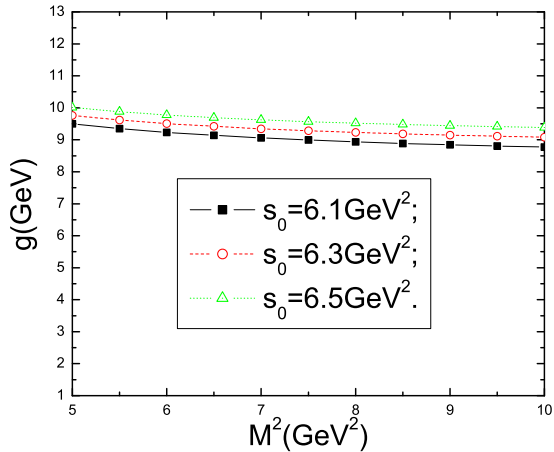}
 \caption{The   $g_{D_{s0}DK}$ with the parameters $M^2$ and $s_0$ . }
\end{figure}

\begin{figure}
 \centering
 \includegraphics[totalheight=7cm]{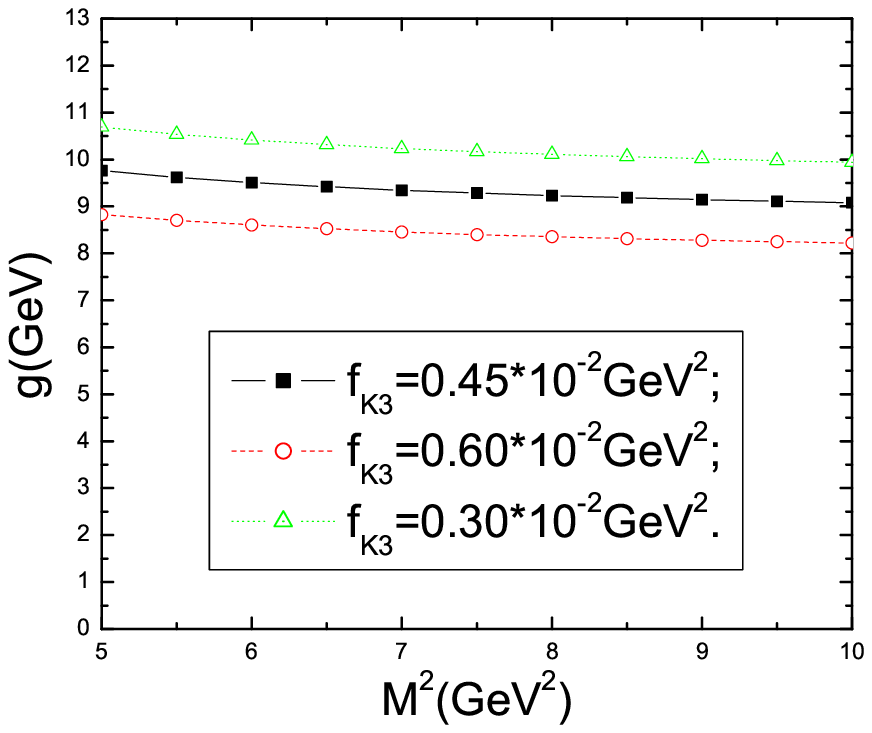}
 \caption{The   $g_{D_{s0}DK}$ with the parameters $M^2$ and $f_{K3}$ . }
\end{figure}

\begin{figure}
 \centering
 \includegraphics[totalheight=7cm]{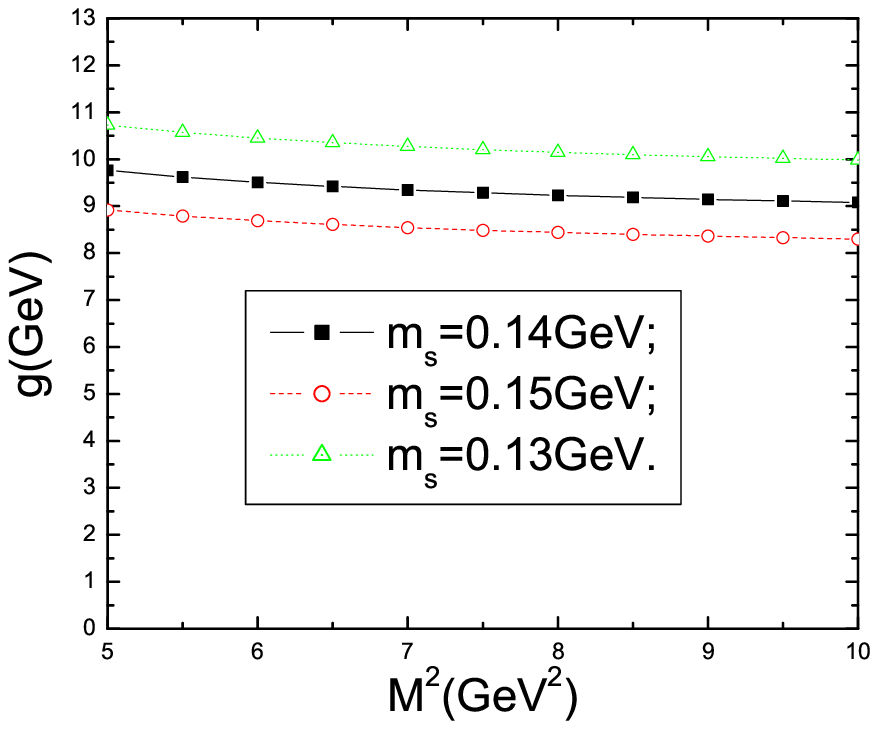}
 \caption{The   $g_{D_{s0}DK}$ with the parameters $M^2$ and $m_s$ . }
\end{figure}

\begin{figure}
 \centering
 \includegraphics[totalheight=7cm]{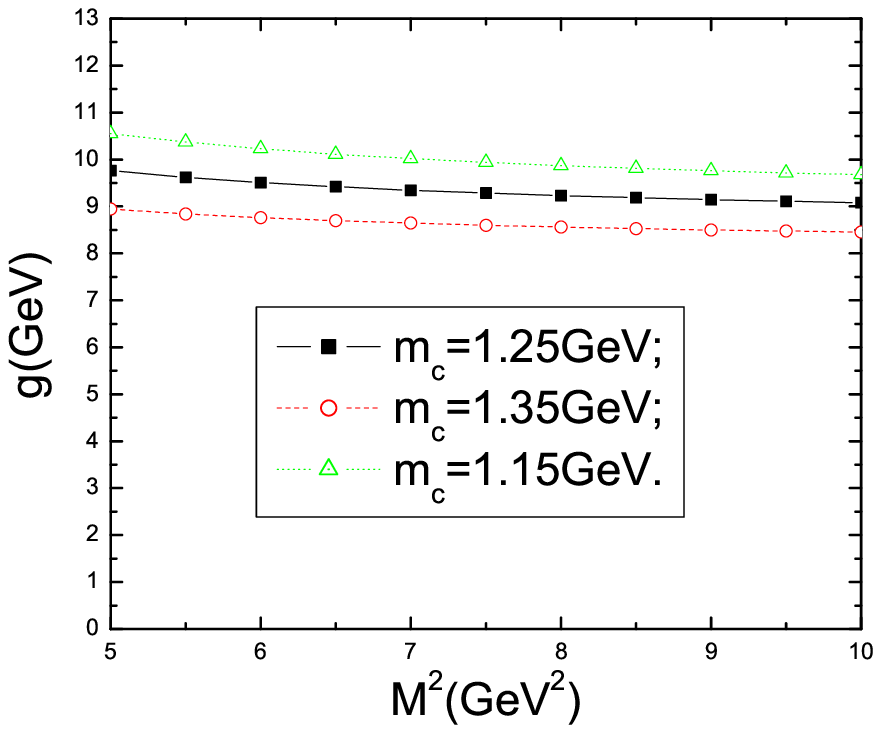}
 \caption{The   $g_{D_{s0}DK}$ with the parameters $M^2$ and $m_c$ . }
\end{figure}

\begin{figure}
 \centering
 \includegraphics[totalheight=7cm]{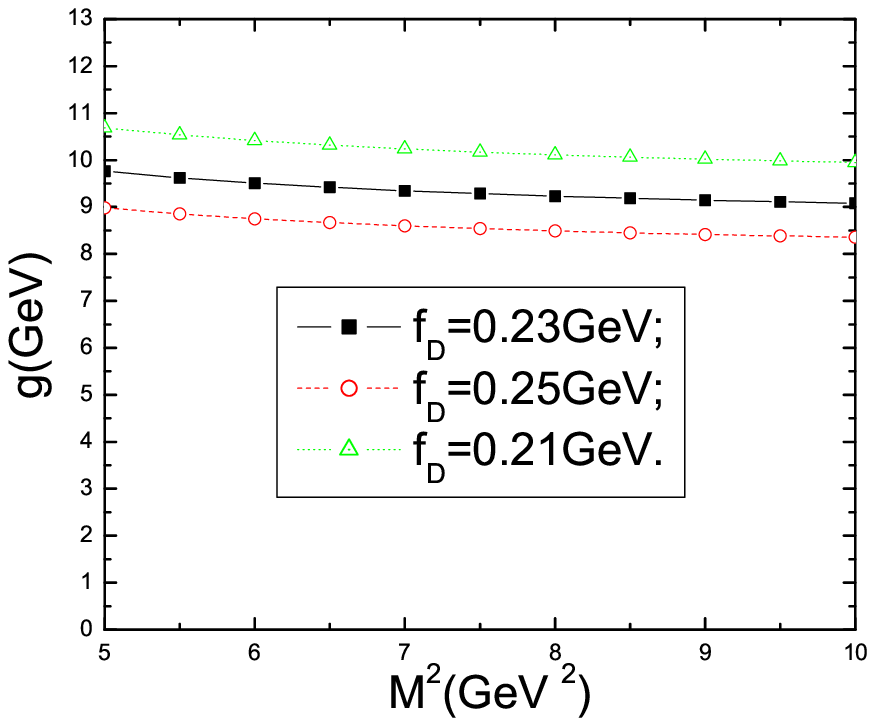}
 \caption{The   $g_{D_{s0}DK}$ with the parameter $M^2$ and $f_{D}$ . }
\end{figure}

\begin{figure}
 \centering
 \includegraphics[totalheight=7cm]{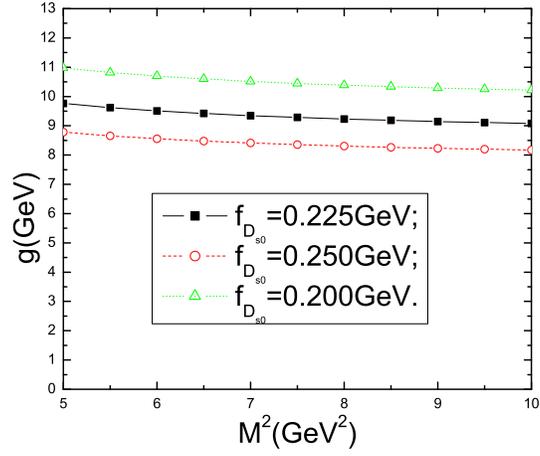}
 \caption{The   $g_{D_{s0}DK}$ with the parameters $M^2$ and $f_{D_{s0}}$ . }
\end{figure}

\begin{figure}
 \centering
 \includegraphics[totalheight=7cm]{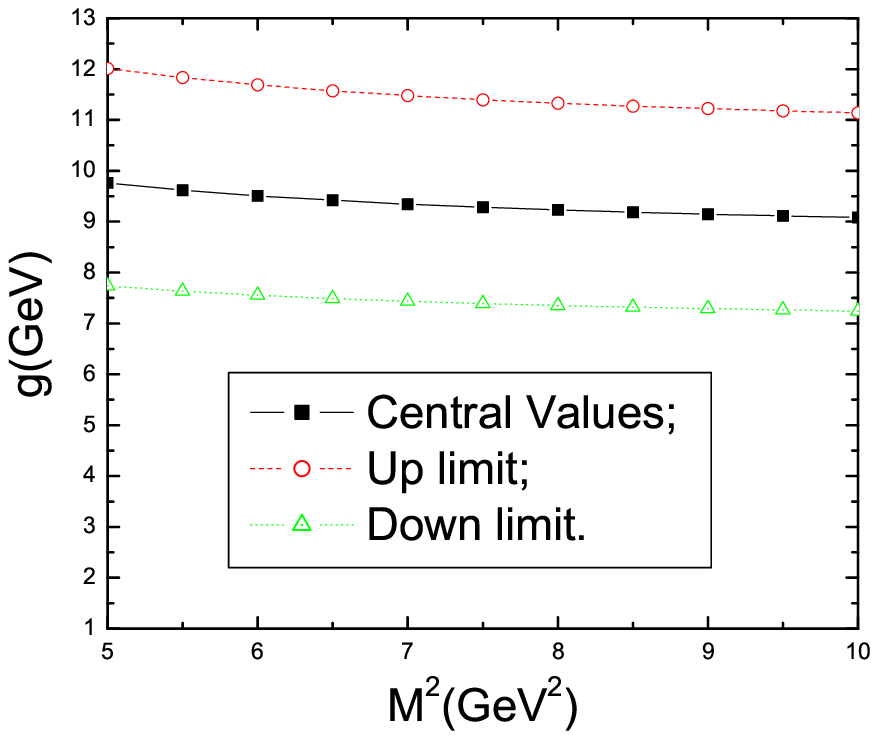}
 \caption{The   $g_{D_{s0}DK}$ with the parameter $M^2$. }
\end{figure}
\section{Conclusions}

In this article, we take the point of view that the  charmed scalar
meson $D_{s0}(2317)$ is the conventional $c\bar{s}$ meson and
calculate the strong coupling constant $g_{D_{s0} D K}$ within the
framework of the light-cone QCD sum rules approach. The  numerical
values for the scalar-$DK$ coupling constant $g_{D_{s0} D K}$ are
compatible with the existing estimations although somewhat larger,
the large values support the hadronic dressing mechanism. Just like
the scalar mesons $f_0(980)$ and $a_0(980)$, the scalar meson
$D_{s0}(2317)$ may have small $c\bar{s}$ kernel  of typical
$c\bar{s}$ meson size. The strong coupling to virtual intermediate
hadronic states (or the virtual mesons loops) can result in  smaller
mass  than the conventional scalar $c\bar{s}$ meson in the
constituent quark models, enrich the pure $c\bar{s}$ state with
other components. The $D_{s0}(2317)$  may spend part (or most part)
of its lifetime as virtual $ D K $ state.

\section*{Acknowledgments}
This  work is supported by National Natural Science Foundation,
Grant Number 10405009,  and Key Program Foundation of NCEPU. The
authors are indebted to Dr. J.He (IHEP), Dr. X.B.Huang (PKU) and Dr.
L.Li (GSCAS) for numerous help, without them, the work would not be
finished.

\end{document}